\documentclass[pr,aps,praps,amsbsy]{revtex4}
\usepackage{graphics}
\begin{document}

\title{\normalsize{\bf{RADIATION OF RELATIVISTIC PARTICLES FOR QUASIPERIODIC MOTION
IN A TRANSPARENT
MEDIUM}}}

\author{ S. Bellucci$^1$ and V.A. Maisheev$^2$\\
 $^{1}$\it{ INFN - Laboratori Nazionali di Frascati, P.O. Box 13, 00044 Frascati, Italy} \\
 $^{2}$\it{ Institute for High Energy Physics, 142281, Protvino, Russia }}

\begin{abstract}
The radiation of  relativistic charged particles for the quasiperiodic motion in
a transparent medium is considered. For motion of the general kind
the differential probability of the process is obtained. For planar
motion the spectral intensity of radiation is found.
The different cases of radiation in the medium-filled undulators are studied.
In particular, the influence of  Cherenkov radiation on
the undulator one is discussed.

PACS number(s): 41.60.-m
\end{abstract}
\maketitle
\section{Introduction}

At the present time  such power sources of  x-rays as undulators \cite{BL}
are widely used in various fields of science. In a number of
papers \cite{Ka,Bar,HI,Bo,De,Ko,MU,RO}, with the aim of increasing
of the energy of emitted
photons, the crystal undulator was considered.
The recent paper\cite{KSG}  contains a rather complete list of references related
to various problem of crystal undulators \cite{fn}.

 One of  the proposed
constructions \cite{SB, SB1} was created and tested in a positron beam.
Preliminary results of the experiment \cite{Che} give an indication on
the observation of  undulator radiation.  Calculations of the
expected intensity for this experiment were
based on the theory \cite{BKS} of radiation for quasiperiodic motion
in vacuum.
This theory allows one to perform the calculation of  radiation spectra
for motion of the general type and different parameters of the undulator.
In calculations with our experimental conditions we use
the theory \cite{BKS} in the frame of
the classical electrodynamics.

However, in the recent papers \cite{RO1,RO2} the possibility of
an appreciable influence of the medium polarization
on the spectral intensity in crystal undulators was shown.
In Ref. \cite{RO1} the process was considered only in the dipole approximation.
In  Ref. \cite{RO2} this process was studied for a specific construction
of the undulator and hence for specific trajectories of particles.
In  both cases the radiation of the first
harmonic  was considered.

In this paper we want to extend the theory \cite{BKS} to the
case of a transparent medium. We will study the pointed process
in the general case of dielectric function $\varepsilon(\omega)$ (where $\omega$
is the frequency or energy of the emitted photon), which may be larger or smaller
than 1. The arising phenomena in  different cases will be shortly discussed.
In this paper we employ units such that $\hbar= c=1$.

Note that a large number of problems of radiation for  charged particles
moving in various media were considered in Ref.\cite{GVL}.
Here the investigations of the radiation in medium-filled undulators
are also presented. However, these results concern mainly
the total radiation intensity.
We  also point out the paper \cite{GK} where the radiation for
quasiperiodic motion was studied in a wide range of undulator
parameters.

\section{Radiation energy losses of particle}
The well-known formula \cite{JD,TM} for  the radiation energy
losses of  a moving particle
takes into account the dielectric function of the medium.
The analogous formula in \cite{BKS} differs from the above pointed  one
and was obtained for the vacuum. With the aim of extending of
the theory \cite{BKS} (in the frame of the classical electrodynamics)
to the case of a transparent medium we should find its corresponding
representation.

The Fourier transform of the vector potential for the electromagnetic field
of a charged particle moving in the isotropic transparent medium
has the following form  \cite{TM}:
\begin{equation}
{\bf{A}}(\omega, r)=
e {\exp{[ik r]} \over r}
\int {\bf{v}}(t) \exp \{i[\omega t - {\bf{k r_0}}(t)]\}dt\,,
\end{equation}
where ${\bf{k}}= \sqrt{\varepsilon}\omega {\bf{n}}$, $\varepsilon (\omega)$
is the dielectric function ($\varepsilon$ is a real positive value),
${\bf{n}}$ is the unit vector in the direction of the photon motion,
${\bf{v}},{\bf{r}}_0 $ are the particle velocity and its radius-vector,
 $r$ is the distance from the point, where the particle
(with the charge e) is located in the moment of time $t$. This relation
is valid for large $r$.

Using this equation we find (analogously to \cite{BKS})
the magnetic and electric (${\bf{E}}$)  fields.
The radiated energy $d{\cal{E}}({\bf{n}}, \omega)$ in an elementary
 solid angle $d\Omega$ and a frequency range
$\omega, \omega+d\omega$ for the whole time of the process is \cite{LL}
\begin{equation}
d{\cal{E}}({\bf{n}}, \omega)=
\sqrt{\varepsilon} |{\bf{E}}(\omega)|^2 (d\omega/4\pi^2)d\Omega r^2.
\end{equation}

Finally,
we obtain
\begin{eqnarray}
d{\cal{E}}({\bf{n}}, \omega)=e^2 \sqrt{\varepsilon} \int \int
[({\bf{v}}(t_1){\bf{v}}(t_2) - 1/\varepsilon]
\times \nonumber \\
\times \exp\{i[\omega(t_1-t_2)-
  {\bf{k}}[{\bf{r_0}}(t_1)-{\bf{r_0}}(t_2)]]\}
{\omega^2 d\omega d\Omega \over (2\pi)^2} dt_1 dt_2 .
\end{eqnarray}
This equation describes the differential radiation energy losses
of the relativistic particle moving in a transparent medium.
At $\sqrt{\varepsilon} = 1$ Eq.(3) is the same as in \cite{BKS}.

The  relation obtained here contains the peculiarities of  radiation
processes in a medium. Let us calculate for demonstration the radiation
of the relativistic
charge particle moving in the transparent medium with a constant velocity
(Cherenkov radiation). It is easy to take the integrals over $t_1$ and $t_2$
\begin{equation}
d{\cal{E}}({\bf{n}}, \omega)=
e^2 \sqrt{\varepsilon}(v^2-1/\varepsilon) T_m
\delta(\omega-\sqrt{\varepsilon}\omega v \cos\theta)
{\omega^2 d\omega d\Omega \over 2\pi},
\end{equation}
where $T_m$ is the time of a particle motion and $\theta$ is the
polar angle, which determined by the pair of vectors: ${\bf{v}}$ and ${\bf{n}}$.

From here, we get the intensity of radiation per  unit time
\begin{equation}
{d{\cal{E}}(\omega)\over T_m } =
e^2 v (1- 1/(\varepsilon v^2))\vartheta(1- 1/(\varepsilon v^2)) \omega d\omega ,
\end{equation}
where $\vartheta(x)= 1$ at $x > 0$ and  $\vartheta(x)= 0$ at $x < 0$.

\section {Intensity of radiation for quasiperiodic motion}

Let us suggest  that the particle performs a quasiperiodic motion with
the period equal to $T$.

The periodicity of motion allows us to transform
the integral in Eq.(3) into a Fourier series.
In accordance with \cite{BKS} (conserving the notation of the variables)
 we can write
\begin{equation}
\int_{-\infty}^{\infty} v^{\mu}(t) e^{ikx(t)}dt=
v^{T_{\mu}} \sum_{m=-\infty}^{\infty} e^{im\varphi_0}=
2\pi v^{T_{\mu}} \sum_{n}\delta (\varphi_0- 2\pi n),
\end{equation}
where
\begin{equation}
kx(t)=\omega t - \sqrt{\varepsilon} \omega {\bf{nr}}_0(t), \,\,
\varphi_0 = \omega T (1-\sqrt{\varepsilon}{\bf{n}}{\bf{V}}), \, \,
v^{T_{\mu}}=\int_{0}^{T} v^{\mu}(t) e^{ikx(t)}dt.
\end{equation}
Here $v^{\mu}=(1, {\bf{v}}) ,(\mu=0-3)$ is the 4-vector of the particle velocity and
${\bf{V}}= (1/T)\int_0^T{\bf{v}}(t)dt $ is the mean particle velocity.
One can find the mean velocity $V$ and longitudinal velocity $v_\parallel$
\begin{equation}
V \approx 1- (1+ \overline{v_\bot^2}\gamma^2)/(2\gamma^2),\,\,\,
v_\parallel \approx 1 -(1+ v_\bot^2\gamma^2)/(2\gamma^2),
\end{equation}
where $\overline{v_\bot^2}$ is the mean square of the transversal velocity $v_\bot$.
Substituting Eq.(6) into Eq.(3) and calculating the intensity
of  radiation (per unit of time) one can get
\begin{equation}
dI=e^2\sqrt{\varepsilon} {{\omega^2d\omega d\Omega }\over 4\pi T } \sum_{n=-\infty}^{\infty}
\delta(\varphi_0-2\pi n) \{2(|{\bf{v}}^T|^2-|v^{T_0}|^2/\varepsilon)\}.
\end{equation}
One can see that in the general case the number $n$ may be positive as well
as negative (see for an explanation, for example \cite{GVL, GK}).

Below we will find the relativistic relations with
an accuracy up to $\gamma^{-2}$ terms.
Besides, we will obtain the intensity of radiation for any positive
value of $\sqrt{\varepsilon} = 1+(\sqrt{\varepsilon}-1)= 1+ \chi $.
Now we find the following relations:
\begin{equation}
f(t)=\omega t - \sqrt{\varepsilon} {\bf{nr_0}}\omega=
-\chi \omega t + \sqrt{\varepsilon} \omega
[{\theta^2 t\over 2}+ {t \over 2\gamma^2} +\int_{0}^t v_\bot^2 dt
-{\bf{n_\bot x_\bot}}],
\end{equation}
\begin{equation}
\varphi_0/T = \omega (-\chi+ \sqrt{\varepsilon}{(1+ \overline{v_\bot^2} \gamma^2) \over 2\gamma^2}+
{\sqrt{\varepsilon} \theta^2 \over 2})= n \omega_0.
\end{equation}
Then we can find the spectral-angular distribution of the radiation
\begin{eqnarray}
dI= e^2 \sqrt{\varepsilon}{{\omega^2d\omega d\Omega } \over (2\pi)^2} {1 \over \omega_0 \gamma^2}
\sum_{n=-\infty}^{\infty} \delta(\varphi_0-2\pi n)
[\{{(\varepsilon -1)\over \varepsilon} \gamma^2+ {1\over 4\gamma^2} -1\}|I_0|^2 + \gamma^2(|{\bf{I_\bot}}|^2- ReI^*_0 I_{\parallel})],
 \\
I_0=\int_0^{2\pi} e^{if(\psi)}d\psi, \,\,\,
{\bf{I_\bot}}=\int_0^{2\pi} {\bf{v_\bot}}(\psi)e^{if(\psi)}d\psi,
\\
I_{\parallel}=\int_0^{2\pi}{\bf{v_\bot}}^2(\psi)e^{if(\psi)}d\psi,\,\,\, \psi =\omega_0 t,
\\
f(\psi)=n \psi + \omega\sqrt{\varepsilon} \Delta(\psi)/(2\omega_0) -
\omega \sqrt{\varepsilon} {\bf{nx}}_\bot(\psi),
\end{eqnarray}
where
\begin{equation}
\Delta(t)=\omega_0 \int_0^{t}({\bf{v}}^2_\bot (t^{'})- \overline{{\bf{v}}^2_\bot})dt^{'}.
\end{equation}
Here we use Eqs.(10),(11) for finding of $f(\psi)$-function. The necessity of keeping
the term  $1/(4\gamma^2)$ in Eq.(12) will be clear below.
The photon energy and the emission angle can be obtained
from the following relations:
\begin{equation}
\omega= {2\gamma^2n \omega_0 \over
\sqrt{\varepsilon}
 (1+\gamma^2\theta^2 +\rho/2- 2\chi \gamma^2 /\sqrt{\varepsilon})},
\end{equation}
\begin{equation}
\theta^2= {1\over \gamma^2} ({2\gamma^2 n \omega_0 \over \sqrt{\varepsilon} \omega}+
{2\chi \gamma^2 \over \sqrt{\varepsilon}} -1 -\rho/2),
\end{equation}
where $\rho = 2 \gamma^2 \overline{v_\bot^2}$.
Obviously, these relations are not independent and we write them for
convenience of the further discussion.

Eqs.(12)-(15) describe the spectral-angular distribution  of
the relativistic particle radiation  for the quasiperiodic motion in the
transparent and isotropic medium. The trajectory of
the particle are represented in these equations in a general form.
Eqs.(12)-(15) allow us to calculate the spectral (integrated over angular variables)
intensity, with the help of numerical methods, for any particle motion.
However,  for some general enough cases the integrals in Eqs.(13)-(15)
may be taken over
angular variables so as, for example, in the important case of the planar motion.
For the planar motion
\begin{equation}
{\bf{nx}_\bot}= \theta \cos(\varphi)\int_0^t v_\bot(t^{'})dt^{'},
\end{equation}
where $\varphi$ is the azimuthal angle.
After integration over $\theta$ we get
\begin{equation}
dI_p = e^2{\omega d\omega d\varphi \over (2\pi)^3\gamma^2}
\sum_{n=-\infty}^{\infty}
\vartheta (\theta^2)
(\{{(\varepsilon -1)\over \varepsilon} \gamma^2 +{1\over 4\gamma^2} -1\}I_0^2+\gamma^2(I^2_x-I_0I_\parallel)).
\end{equation}
From here on, the term $\vartheta (\theta^2)$ in the sum means
that the function $\vartheta$ (which was defined after Eq.(5)) is equal to
0 or 1 in accordance with Eq.(18).

One can integrate this relation over $\varphi$ and
obtain the following equation for the spectral intensity:
\begin{eqnarray}
{dI\over d\omega}=  -\,{e^2\omega  \over (2\pi \gamma)^2}
\sum_{n=-\infty}^{\infty} \vartheta (\theta^2)
\int_{-\pi}^{\pi}dt_1dt_2
J_0 \left (2\sqrt{\varepsilon} \int_{t_2}^{t_1}d\psi g(\psi)
\sqrt{\xi(n/\sqrt{\varepsilon} -\xi(1+\rho/2 -2\chi \gamma^2/\sqrt{\varepsilon} ))}\right) \times
 \nonumber \\
  \times (1 - \{{(\varepsilon -1)\over \varepsilon} \gamma^2
+{1\over 4\gamma^2} \}+ 1/2(g(t_2)-g(t_1))^2)
 \cos{ ((n-\sqrt{\varepsilon}  \xi \rho/2)(t_1-t_2)
+\sqrt{ \varepsilon } \xi \int_{t_2}^{t_1}g^2(\psi)d\psi) },
\end{eqnarray}
where  $g(\psi)= \gamma[v_x(\psi)-<v_x>]$, $v_x=v_\bot$,  $<v_x>$ is the mean transversal velocity,
$\xi= \omega /(2\gamma^2\omega_0)$ and $J_0(x)$ is the Bessel function.
Then we can get from Eq.(21) the following relation,
in the dipole approximation:
\begin{equation}
{dI\over d\omega}= e^2 \omega
\sum_{n=-\infty,\ne 0}^{\infty} \vartheta (\theta^2)
|x_n|^2
\{n^2-2[\varepsilon -(\varepsilon -1)\gamma^2]
[\xi(n/\sqrt{\varepsilon} -\xi(1+\rho/2 - 2\chi \gamma^2/\sqrt{\varepsilon}))]\} ,
\end{equation}
where $x_n =(1/(2\pi) \int_{-\pi}^{\pi} x(\psi) \exp(-i n\psi)d\psi$ is the Fourier component
of the value $x(t)= 1/\gamma \int_0^t g(\psi)d\psi$
($x(t)/\omega_0$ is the transversal coordinate).
This equation was obtained for the two conditions:
$\varepsilon \omega^2\theta^2 \rho/\gamma^2 \ll 1$ and $\rho \sqrt{\varepsilon}  \xi\ll 1$.
The first condition is the requirement of the smallness of the argument in the
Bessel function,  and the second one means that the cosine in Eq.(21) is
approximately equal to $cos(n(t_1-t_2))$. In spite of the fact that $\rho\ll 1$
the value $ 2 \chi \gamma^2$ may be large enough and hence
the argument of the Bessel function can be also large. Because of this,
the first condition  is also necessary.
At $\varepsilon =1$ this equation has the form
of the well-known dipole approximation ( when $\rho <<1$).

In the case when
\begin{equation}
2 \chi \gamma^2/\sqrt{\varepsilon} > 1+ \rho/2
\end{equation}
the following term (n=0) should be added in Eq.(22):
\begin{equation}
{dI_{n=0} \over d\omega}(\omega)= e^2 \omega
 \{({\varepsilon -1 \over \varepsilon} + {1\over 4 \gamma^4} -{1+\rho/2\over \gamma^2})
-({(\varepsilon-1)\gamma^2 \over \varepsilon}+{1 \over 4\gamma^2}-1)
[2\varepsilon \xi^2(2\chi \gamma^2\sqrt{\varepsilon}-(1+\rho/2))] \overline{ X^2}\},
\end{equation}
where $\overline{ X^2}$ is the mean square of the function
$x(t)$ 

The Eqs.(21)-(24) obtained here are sufficient for the calculation of the spectral
intensity of the relativistic particle for its planar quasiperiodic motion
of the general kind in transparent media. In these equations the
knowledge of the function $\varepsilon$ is required for every computed
photon energy. In particular, the process of calculation (at fixed $\omega$)
consists in testing the
relation $\vartheta (\theta^2) $ for every n (in
the interval $-\infty, +\infty$). For such a testing  Eq.(18) should be used.
However, it is easy to see that at the condition $2 \chi \gamma^2/\sqrt{\varepsilon} < 1+ \rho/2$
only positive numbers $n$ are possible.

\section{Examples of calculations}
In this section we point out the basic peculiarities of the radiation
for quasiperiodic particle motion in the medium. For the detail description
of this process the knowledge of explicit form of the dielectric function is
important.
The aim of our consideration is the application of the equations
obtained in the previous section to the calculations of radiation processes in a transparent
medium.
Note that many peculiarities of similar processes were discussed
in earlier papers \cite{RO1,RO2,GVL,GK,BG,MC,Bar2,BZh,XA}

In the general case the relations obtained here for radiation
in  a medium are valid,
at  the condition of a small influence of this media
on the quasiperiodic particle motion.
The different processes (multiple scattering, ionization energy losses
and others) can modify the motion of particles
and they should be investigated separately.
Various examples of consideration of this problem can be found in the
literature \cite{BKS,TM, BM}. One can assert
that, in the case of small enough values of
$|  \varepsilon-1|$, the influence of the medium on the motion will be insignificant,
but  in every specific case such a possibility should be studied.
Thus, we think that in most, if not all, of the practically
important cases  one has $|  \varepsilon-1| \ll 1$.

It is a well-known fact that  the transparent medium is an idealized substance.
We assume that
a good model of the transparent media is
the media  in which $\varepsilon^" \ll |\varepsilon -1|$,
where $\varepsilon^" $ is the imaginary part of  the dielectric function.

In the general case the numbers of harmonic $n$, which may be radiated,
lie in the range ($- \infty, \infty$). In the vacuum $n \ge 1$, always.
However, at the condition $2 \chi \gamma^2/\sqrt{\varepsilon} < 1+ \rho/2$
all the solved numbers are positive. It is easy to see
from Eqs.(32) and (33) that the
condition $2 \chi \gamma^2/\sqrt{\varepsilon} = 1+ \rho/2$
is practically equal to the threshold of the Cherenkov radiation.
From Eq.(33) it follows that, at the decreasing
of the amplitude of the transversal
motion ($\rho \rightarrow 0$), this equation describes the intensity of
the Cherenkov radiation.
Besides, $dI(n=0) /d\omega =0$ at $2 \chi \gamma^2/\sqrt{\varepsilon} = 1$.
For a demonstration of this fact, the term $1/(4\gamma^2)$ was kept
in Eq.(12). At $\rho \rightarrow 0$ all the remaining terms
($n\ne 0$) in Eqs.(21) and (22) are set to zero.

Let us recall that $\rho=2\gamma^2 \overline{v_\bot^2}$ and hence
we can find the threshold value of the Lorentz factor for Cherenkov
radiation in the general case
\begin{equation}
\gamma^2_{th}= {1 \over 2\chi/\sqrt{\varepsilon}-  \overline{v_\bot^2}}.
\end{equation}
From here, we see that $\gamma_{th}$ is increased at the increasing of the mean
squared transversal velocity.
We also see that, for allowing the possibility of Cherenkov radiation, the
realization of the condition $ \overline{v_\bot^2} < 2 \chi/\sqrt{\varepsilon}$
is necessary.

Let us consider Eqs.(17) and (18). We see that for Cherenkov radiation ($n=0$)
\begin{equation}
\theta^2_{Ch}= {1 \over \gamma^2} ({2\chi \gamma^2 \over \sqrt{\varepsilon}}
-1-\rho/2).
\end{equation}
This result shows that the angle of Cherenkov radiation also depends
on the $\rho$-value. From Eq.(17) we see that no limitations on
the energy of the emitted photons (for $n=0$).

Now we consider the case of the usual amorphous media.
At high enough frequencies
of  photons the dielectric function has the following simple form:
\begin{equation}
\varepsilon = 1 -{\Omega^2_p \over \omega^2}.
\end{equation}
where $\Omega^2_p=4\pi n_e e^2/m_e$ , $n_e$ is the electron density and
$m_e$ is the electron mass.
Substituting this relation in Eq.(27)  we
obtain  approximately (at the condition  $ \Omega_p/\omega \ll 1 $)
for radiation of the $n$-th harmonic
\begin{equation}
(1+\rho/2)\omega^2 - 2\gamma^2n \omega_0 \omega + \gamma^2\Omega^2_p \le 0.
\end{equation}
It means that radiation (of $n$-th harmonic) is possible when $\gamma n \omega_0 > \Omega_p \sqrt{1+\rho/2}$
and the resolved photons energies lie in the interval  $\omega_{-} \le \omega  \le \omega_+$, where
\begin{equation}
\omega_{\pm}={\gamma^2 n \omega_0 \pm \sqrt{\gamma^4 n^2\omega_0^2- \Omega^2_p \gamma^2(1+\rho/2)} \over 1+\rho/2}.
\end{equation}
The threshold value of the Lorentz  factor for the harmonic number $n$  is equal to
\begin{equation}
\gamma_{th}={\Omega_p \over \sqrt{n^2 \omega^2_0-\overline{v^2_\bot}\Omega_p^2}} \approx
{\Omega_p \over \omega_0 \sqrt{n^2-a^2 \Omega_p^2/2}},
\end{equation}
where $a$ is the amplitude of the particle deflection. It is obvious that
for radiation of the $n$-th harmonic: $a\Omega_p < \sqrt{2} n$.
These results are in agreement with those in \cite{RO1, RO2}( for the 1st
harmonic) and \cite{BZh}.

It is well known that in  usual media the dielectric function is smaller than 1
at high enough photon energies. Thus, the Cherenkov undulator radiation
is possible mainly at the photon energies $< \sim 10 $ eV. Besides, there exists
the possibility to observe this radiation on the
photoeffect absorption edges \cite{BG,MC}. In this case the energy of emitted
photons has a value $ < \sim 1$ KeV.

In accordance with quantum electrodynamics \cite{LL1} the electromagnetic vacuum
represents the medium in which the dielectric function may be larger than 1.
However, for electric fields which may be obtained in laboratories
($< 10^6$ gauss)
the value $\varepsilon -1$  is very small and the particles with Lorentz factors
larger than $10^{10}$ can feel this value.
In paper \cite{MMF} the Cherenkov radiation was predicted in silicon
single crystals (i.e. the analogue of the quantum undulator) for particles
with $\gamma > 10^8$. However, our considerations allow us to predict
the specific radiation of negative harmonics in single crystals.
A similar effect is also applied to the propagation
of high energy charged particles
in power laser waves \cite{MV,Dr}.

Below we present some examples of calculations of the radiation of
relativistic particles for  quasiperiodic motion in the medium. These
calculations were done with the use of Eq.(21) assuming that the
particle motion in the transversal plane is harmonic: $v_\bot = a\omega_0\cos \omega_0 t$.

Figs.1 and 2 illustrate the influence of media on the radiation in
the crystal undulator \cite{SB,SB1,Che}.
 In the silicon single crystal at photon energies
larger than 10 KeV the dielectric function is smaller than 1.
The disappearance of the first harmonic in such media
is shown in Fig.1. Fig.2 illustrates the medium influence in the case
when the first harmonic is partially radiated.
Notice that these figures were made only for aim of illustration and
do not take into account many peculiarities of the real process
(such as the influence of the channeling motion, etc.).

Let us consider the particle radiation in the undulator with
the dielectric function larger than 1. In practice it may be a gas-filled
undulator. Let the energy of particles moving in the undulator
satisfy the condition for Cherenkov radiation (see Eq.(26)).
Then the connection between the angle of radiation
of the $n$-th harmonic and the angle of the Cherenkov radiation follows
from Eq.(18)
\begin{equation}
\theta^2(n)= \theta^2_{Ch} + {2n\omega_0 \over \sqrt{\varepsilon} \omega}.
\end{equation}
From here,  we get that the condition of radiation of the $n$-th harmonic is
\begin{equation}
 n > - \,{\theta_{Ch}^2 \sqrt{\varepsilon} \omega \over 2\omega_0}.
\end{equation}
Obviously all the positive $n$ satisfy this condition and negative $n$
satisfy Eq.(26) starting from some number $n_{min}$.

Let us imagine a medium with constant dielectric function
($\varepsilon > 1$)  for
all photon energies. From our consideration (see Eqs.(25),(26),(31),(32) and the condition
for the Cherenkov radiation to take place) it follows
that in this medium the positive harmonic is radiated at all photon energies.
There is a threshold for negative harmonics in this medium.
In this case
the radiation of these harmonics takes place at all above-threshold energies, and
with the increasing of  photon energy the number of radiated harmonics
also grows. In particular, the frequency $\omega_0$ determines only
the threshold energy of radiated harmonics.
This consideration shows that the character of radiation (at the pointed
conditions) is appreciably different than in the usual undulator.

For illustration of this case we carry out the calculation of the
propagation of the
beam with the Lorentz factor equal to 400
in the gas-filled undulator with a period equal to 10 cm.
We also assume the value $\chi = 10^{-4}$ at photon energies
lower than 1 eV, and $\chi = 0$ at energies higher than 1 eV.
This value is several times smaller
than in many gases at the atmospheric
pressure. The  energy range of photons corresponds approximately to visible light.
Fig.3 illustrates the spectral intensity of the radiation at $\rho =0.39$.
In this case  harmonics with the numbers -1, 0, 1  are predominantly radiated.
We see that at small energies the radiation of the zeroth harmonic
dominates (in accordance with Eq.(24)). The total intensity grows proportionally
to the photon energy and hence is equal to the intensity of Cherenkov radiation
in any medium, which is characterized by the corresponding $\varepsilon$-value.

Fig.4 illustrates the behavior of the intensity of radiation,
depending on the $\rho$-parameter. One can see that at $\rho \approx 62$
the intensities of all the negative and zeroth harmonics disappear.
The structure in curves at large enough $\rho$ reflects the disappearance
of the negative harmonics. The peak at $\rho \approx 38$ corresponds
to  the harmonic with $n= -3$. At the fixed Lorentz factor one has
a threshold value $\rho_{th}=4\chi \gamma^2/\sqrt{\varepsilon}-2$.
Fig.5 shows the angle of radiation of harmonics and the intensity of radiation
at the fixed photon energy and $\rho$-parameter.
From our consideration it follows, firstly, that the undulator
distributes the Cherenkov radiation over its
harmonics, and secondly, that the intensity of radiation in such
a medium is much higher than in the vacuum.

Our results about radiation in
media with $\varepsilon <1$ are in agreement with the main conclusions
of the papers \cite{RO1,RO2}.

The Cherenkov radiation for quasiperiodic motion was studied
in Ref. \cite{GK}. In this paper the radiation process was considered
for specific motion and different undulator parameters. However,
the particular calculations and illustrations for the case $\varepsilon > 1$
and $\rho > 1$  are absent.
The Cherenkov radiation was investigated in more detail at small
undulator parameters.
The conclusion in this paper, i.e. that the undulator radiation is negligible
in comparison with the Cherenkov one, is in agreement with
our results at $\rho < 1$.

\section{Conclusions}
In this paper we considered the radiation process in a transparent medium.
We obtained (on the basis of such a relation for vacuum \cite{BKS})
 the general relation for radiation energy losses of the relativistic
particle.  With the help of this formula we extended the theory \cite{BKS}
of the radiation for the quasiperiodic motion  to the case of a transparent medium.
We got the relations describing the spectral intensity for the case
of the planar motion, which may be prescribed by any analytical equation.
The various possibilities of radiation in transparent media were discussed.

\section{Acknowledgements}
We would like to thank R.O. Avakian for drawing to our attention
the problem of medium for radiation in crystal undulators.
We are grateful of V.G. Baryshevsky and V.V. Tikhomirov for
useful explanations concerning the first papers devoted
to the crystal undulators.

This work was partially  supported
by the Russian Foundation for Basic Research
(grant 05-02-17622
              and  grant 05-02-08085ofi-e.)

\newpage
\section{Figure captions}

\hspace{0 mm} Fig.1 Intensity of radiation in
a silicon crystal undulator with the
period and amplitude  equal to 0.05 cm and 100 angstrom, respectively.
The energy of the positron beam is 10 GeV. Thin and thick curves
correspond to radiation in vacuum and media, respectively.
The parameter $\rho$ has the value $\rho= 6.4$.

Fig. 2. The same as in Fig.1, but  with  the amplitude $a= 55$ angstrom
and $\rho = 1.83$.

Fig.3 Intensity of radiation in gas-filled undulator
as a function of the photon energy.
Curves -1, 0, 1 correspond to radiated harmonics
with $n=$ -1, 0, 1. The thick curve is the total intensity.
The dotted curve is the intensity
in vacuum enlarged 500 times (with the values of the other parameters remaining unchanged).

Fig.4 Intensity of radiation in gas-filled undulator
as a fuction of $\rho$-parameter.
Curve 0 corresponds to the zeroth harmonic, curve 1 (-1)
corresponds to the sum of intensity of all the positive
(negative) harmonics. The thick curve is the total intensity.
The energy of radiated photons is equal to 1 eV.

Fig.5  Radiation of the harmonics in the gas-filled undulator.
The angle $\theta$ is along the abscissa axis and the intensity of radiation
of the $n$-th harmonic is along the ordinate axis.
The numbers above intercepts, showing the intensity, are
the numbers of the harmonics. The intensity of the 6th and 7th harmonics
are invisible (due to their small values).
The energy of radiated photons is equal to 1 eV.
The angle $\theta$ for the 0-th harmonic
is independent of the photon energy and for other harmonics these angles
are changed in accordance with Eq.(31). The parameter $\rho$ takes the value $\rho= 3.8$.

\newpage
\begin{figure}
\scalebox{0.7}{\includegraphics{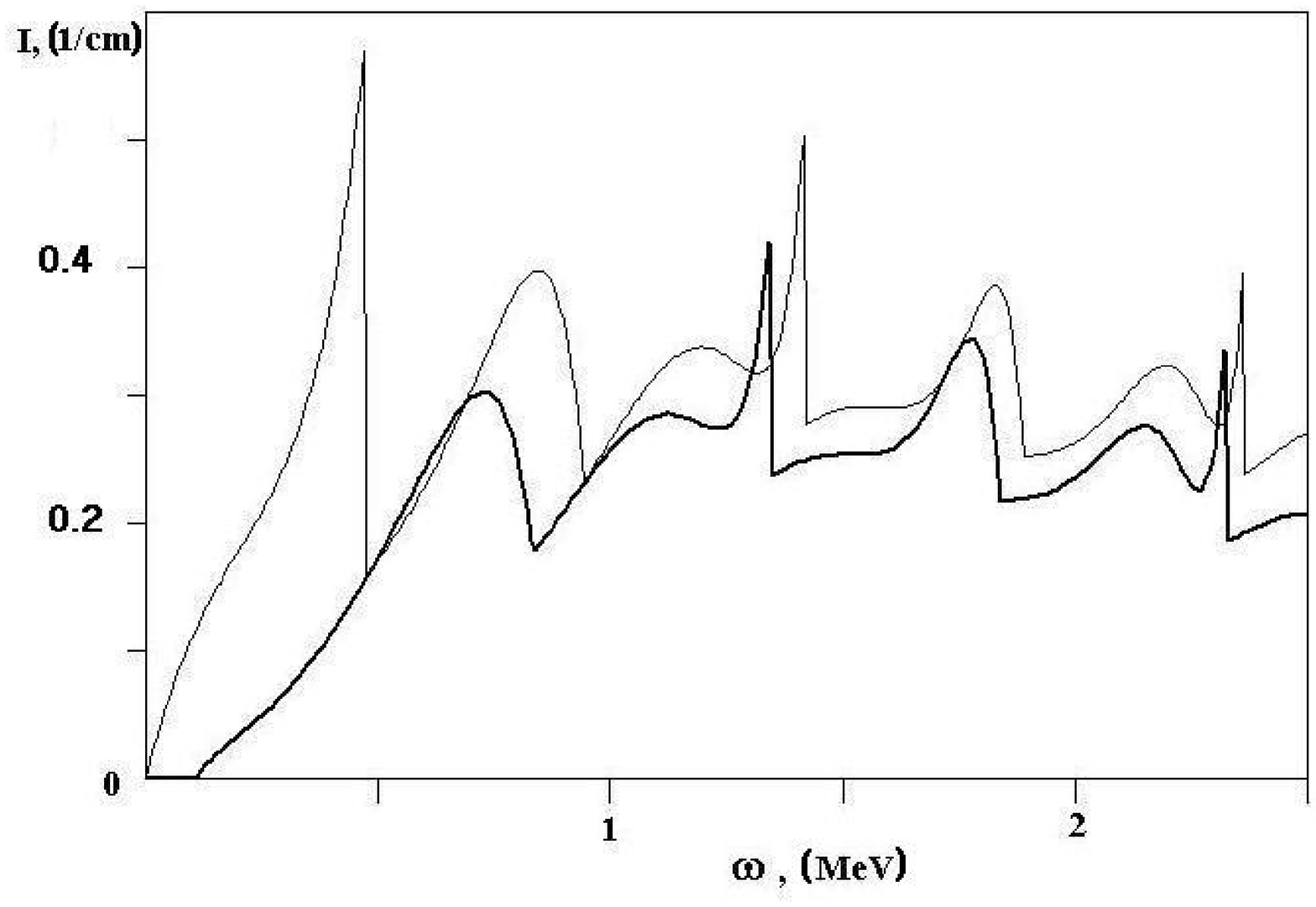}}
{\caption{
              }}
\end{figure}

\vspace*{100 mm}

\begin{figure}
\scalebox{0.7}{\includegraphics{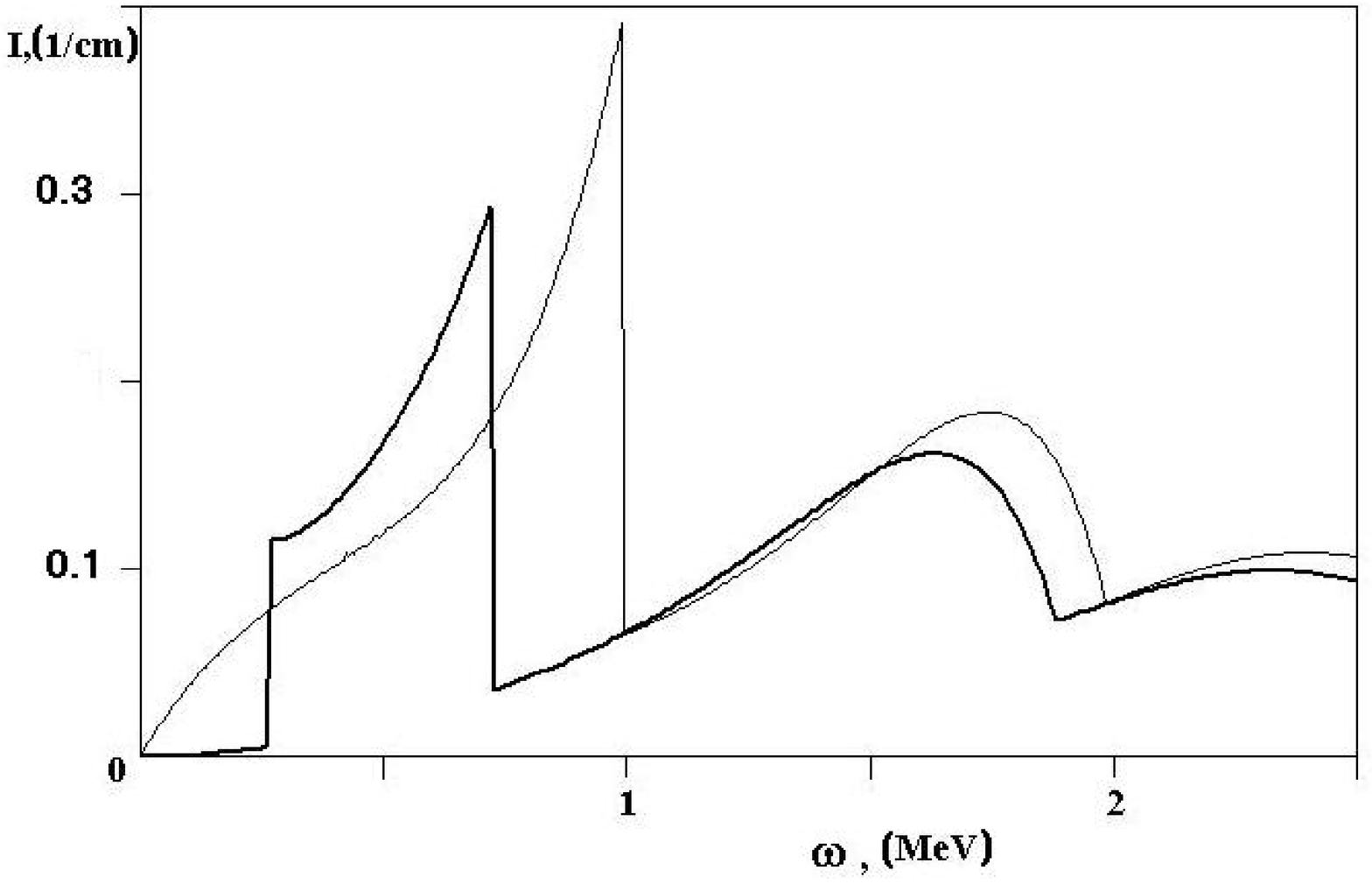}}
{\caption{
              }}
\end{figure}

\vspace*{100 mm}

\begin{figure}
\scalebox{0.7}{\includegraphics{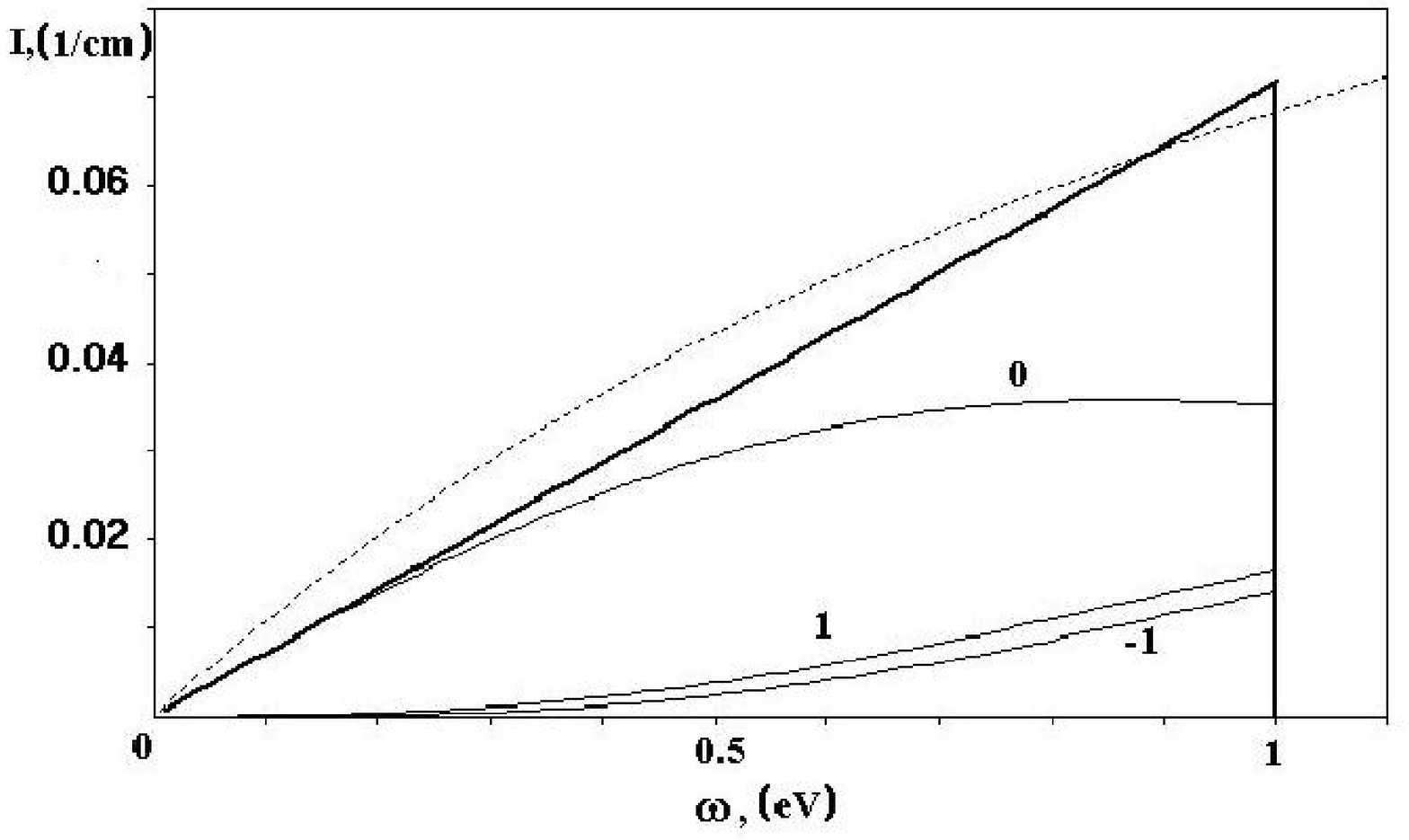}}
{\caption{
              }}
\end{figure}
\vspace*{100 mm}

\begin{figure}
\scalebox{0.7}{\includegraphics{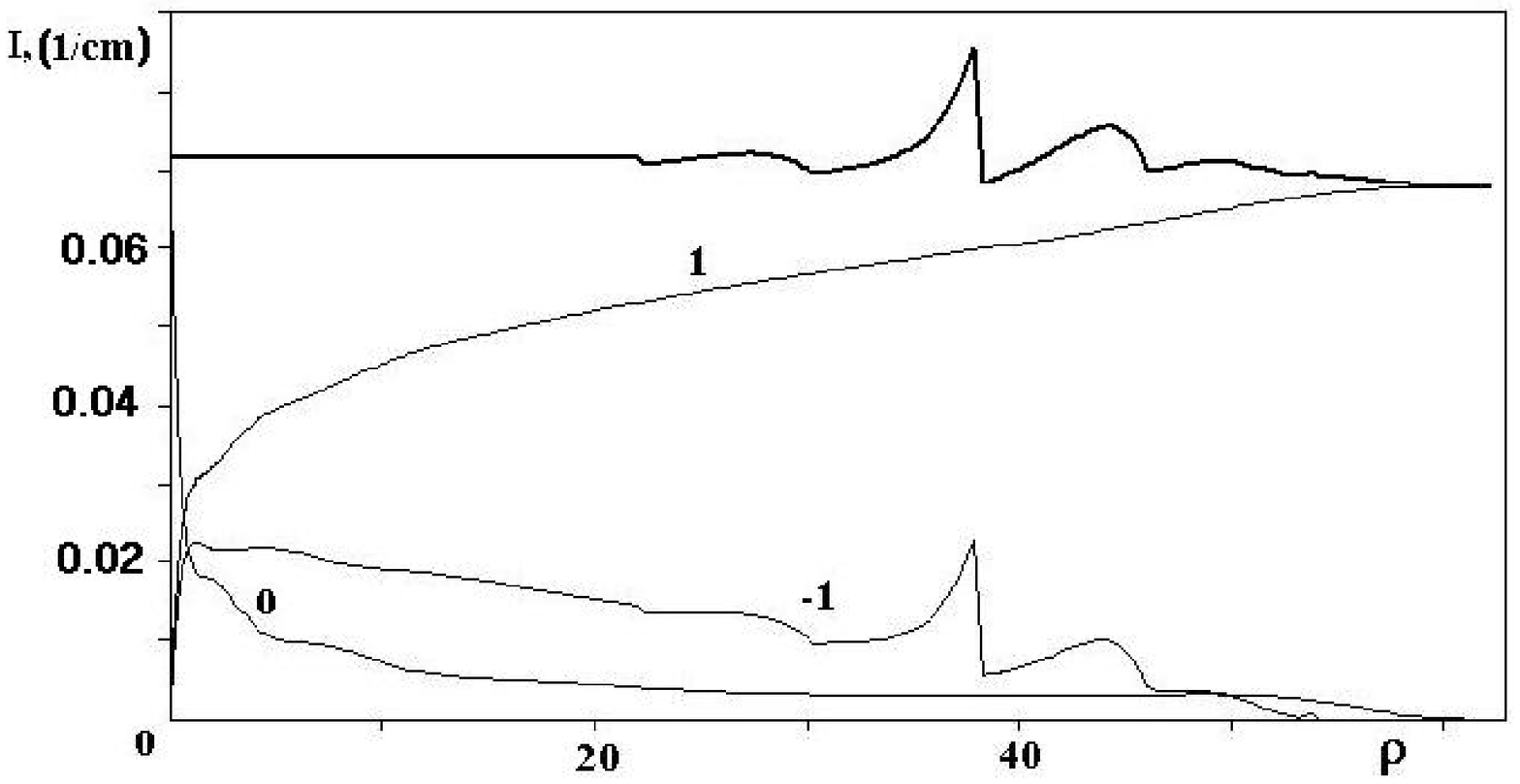}}
{\caption{
              }}
\end{figure}
\vspace*{100 mm}

\newpage
\begin{figure}
\scalebox{0.7}{\includegraphics{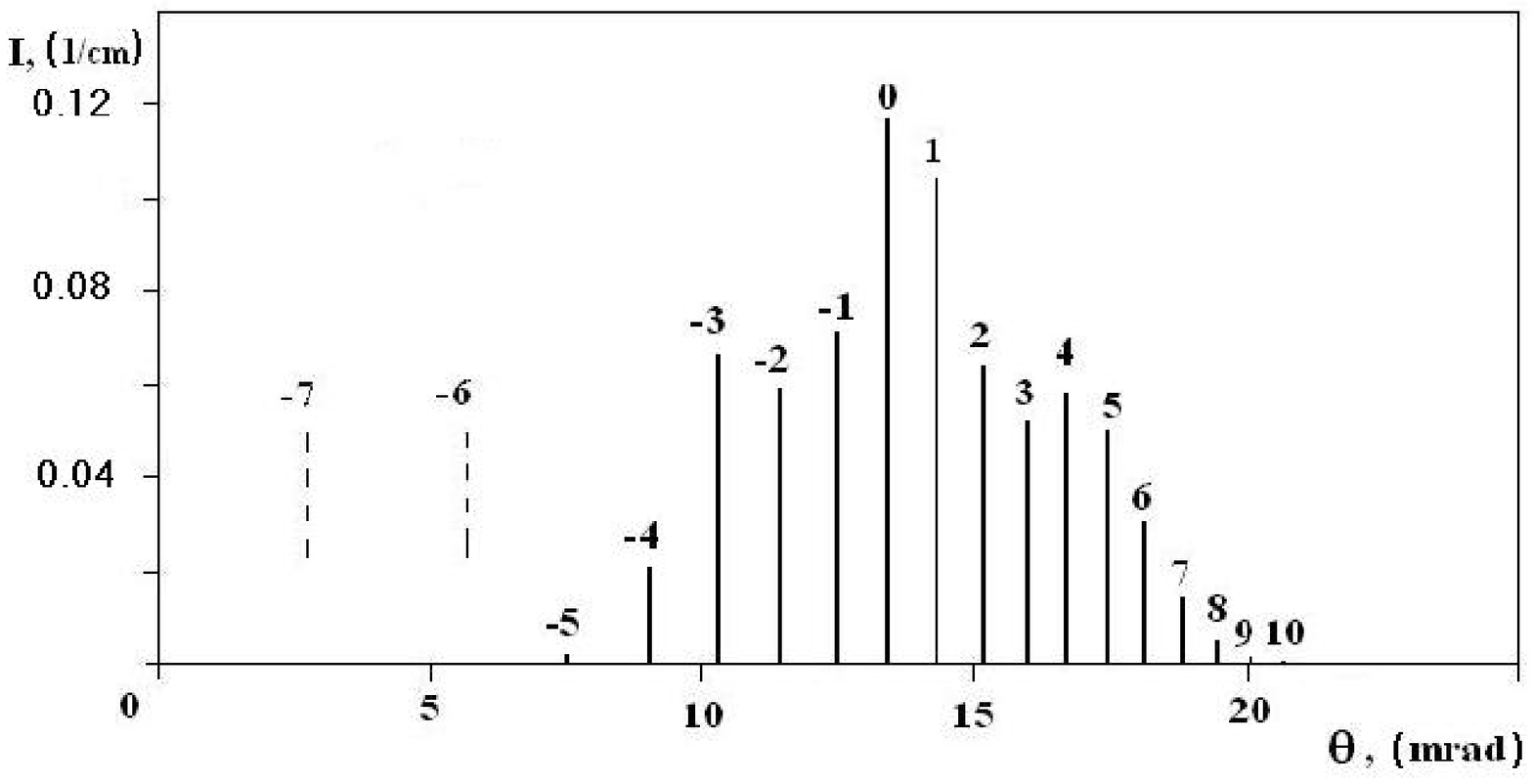}}
{\caption{
              }}
\end{figure}


\begin{thebibliography}{99}
\bibitem{fn}
{We, following the
authors of Ref.\cite{Bar,Bar1}
of one the first publication in this field)   cannot
agree with some statements of the article \cite{KSG} which touch upon
some historical aspects and assessments of the contribution of
different authors to this issue.}
\end{thebibliography}
\end{document}